\documentclass[12pt]{iopart}

\usepackage{graphicx}
\usepackage{units}
\usepackage{hyperref}
\newcommand{\ein}[1]{\, \unit{#1}}

\begin{document}
	
	\title{Trap-loss spectroscopy of Rydberg states in ytterbium}
	
	\author{C. Halter, A. Miethke, C. Sillus, A. Hegde and A. Görlitz}
	
	\address{Institut für Experimentalphysik, Heinrich-Heine-Universität, Düsseldorf, Deutschland}
	\ead{axel.goerlitz@hhu.de}
	\vspace{10pt}
	\begin{indented}
		\item[]October 2022
	\end{indented}

\begin{abstract}
We present  an experimental study of the Rydberg $^1S_0$\,- and $^1P_1$\,-  series of ytterbium for principal quantum numbers in the range of $n=70$ to $90$. The study is performed using trap loss spectroscopy in a magneto-optical trap operating on the $6^1S_0 \rightarrow 6^1P_1$ transition at $399\ein{nm}$. Compared to the  commonly used Rydberg spectroscopy method using field-ionization and ion detection, trap loss spectroscopy is significantly simpler and requires a less sophisticated experimental setup. Using this method we determine relative values of the scalar and tensor electric polarizabilities of both, $6sns\,^1S_0$\,- and $6snp\,^1P_1$\- Rydberg states. 
\end{abstract}
%\keywords{Rydberg, Ytterbium, spectroscopy}
\maketitle
\section{Introduction}

Due to their special properties, Rydberg atoms have attracted the attention of experimental atomic physicists for a long time \cite{Gallagher2004}. In recent years the possibility to produce Rydberg atoms with kinetic temperatures in the $\ein{mK}$\,- or even $\ein{\mu K}$\,- regime has opened up fascinating new possibilities for the precise control of Rydberg atoms in particular to exploit the long-range dipole-dipole interaction \cite{Comparat2010}. This strong interaction may be used to observe a dipole blockade which can lead to a highly-correlated state in a mesoscopic atom cloud \cite{Weimer2008} or to entangle individual atoms \cite{Lukin2001, Urban2009} offering the possibility to use Rydberg atoms for quantum information processing \cite{Saffman2010}.\\

In the majority of experiments with ultracold Rydberg atoms, alkaline atoms such as rubidium have been used, where only one electron can be easily excited to a Rydberg state. For alkaline earth atoms or other atoms with two-electrons in the outermost shell, such as ytterbium (Yb), new possibilities arise since the   atomic core remains optically active if only one electron is excited to a Rydberg state \cite{Bell1991}. For example, for Rydberg states with low angular momentum so-called isolated-core excitation,  of the second electron can lead to auto-ionization of the Rydberg electron. This phenomenon can  be used to control Rydberg excitations \cite{Lehec2021,Burgers2022} or directly image Rydberg atoms without the need to apply high-voltage pulses\cite{Lochead2013, McQuillen2013}. Another attractive application is trapping of Rydberg atoms \cite{Wilson2022}. Very recently, several research groups have started to exploit the particular properties of two-electron Rydberg atoms for applications in quantum technology. The most promising platform consists of tweezer arrays in which atoms can be trapped at single sites and manipulated using transitions to Rydberg states as has recently been demonstrated for strontium \cite{Madjarov2020,Schine2022} and Yb \cite{Ma2022,Okuno2022}.  \\

For all applications using Rydberg atoms, a precise knowledge of the level structure and the specific properties of the Rydberg states is essential. For Yb, several low-resolution studies have been performed in hot atomic samples\cite{Wyart1979,Aymar1980,Camus1980,Xu1994}. Here we present a precise measurement of the binding energies and polarizabilities of the most abundant isotope $^{174}$Yb using simple and efficient trap-loss spectroscopy in a magneto-optical trap. In particular, we determine the binding energies of the $6sns\,^1S_0$\,- and $6snp\,^1P_0$\,- Rydberg series for principal quantum numbers ranging from from $n=70$ to $n = 90$ and measure the relative polarizabilities. Our measurements complement recent results on Yb Rydberg energy levels obtained in significantly more complicated setups \cite{Okuno2022, Lehec2018}.

\section{Experimental Setup}

In our experiment, trap-loss spectroscopy is performed in a standard six-beam {magneto optical trap (MOT) using the isotope $^{174}$Yb. The MOT operates on the broad $6s^2\,^1S_0 \rightarrow 6s6p\, ^1P_1$ transition at $399\ein{nm}$  which is continuously loaded from a Zeeman slower. The laser light for the MOT and the Zeeman slower is provided by a home-built diode laser system in master/slave configuration (Nichia NDHV310ACAEI/Nichia NDV4313E) (see Fig.\,\ref{fig:level_scheme_setup}\,b)). The master laser is frequency-stabilized to a Doppler-free Yb spectroscopy. For precise control of the laser frequencies acousto-optical modulators are used. 

\begin{figure}[h]
\begin{center}	
\includegraphics[width= \textwidth]{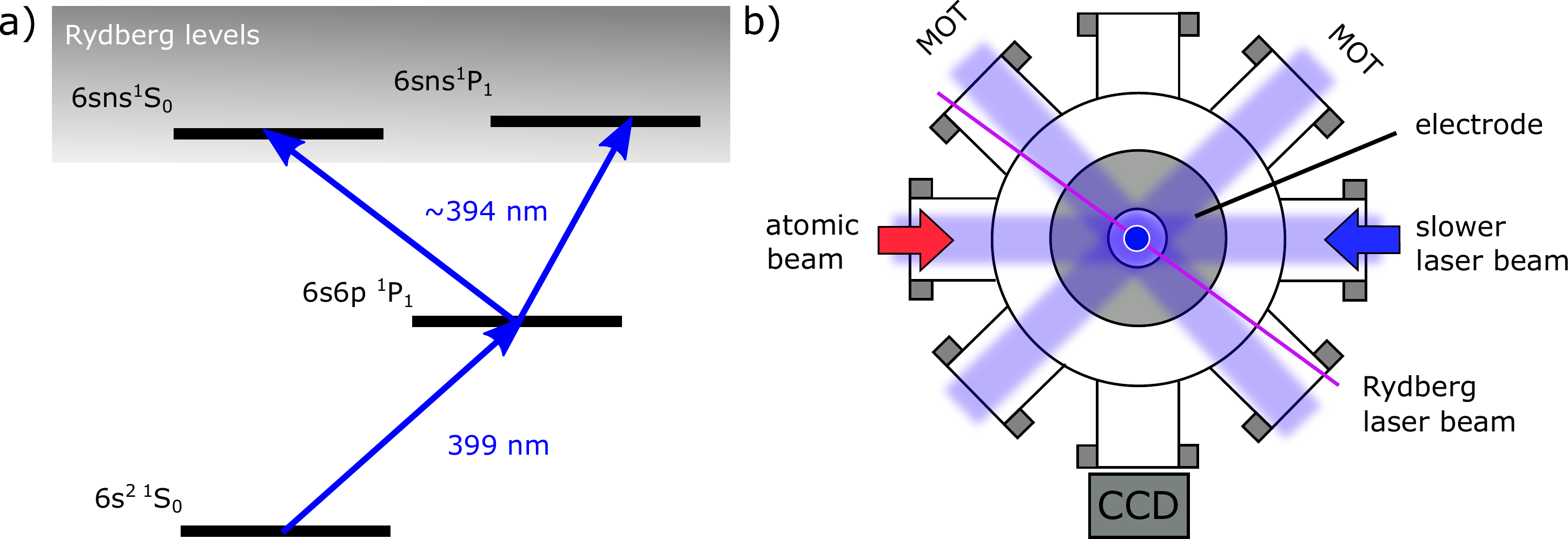}

\end{center}
\caption{a) Relevant level scheme for Rydberg excitation of Yb: The MOT is operated on the $6s^{2}\,^1S_0 \rightarrow 6s6p\,^1P_1$ transition. Rydberg atoms are produced using radiation at $394\ein{nm}$ which connects the excited $6s6p$ $^1P_1$ MOT state to the Rydberg manifold. \\ b) Top view of the experimental setup: The MOT is loaded from a Zeeman slower and operates in a standard six-beam configuration. For Rydberg excitation, the Rydberg laser beam is focused onto the MOT through one of the MOT beam windows. The fluorescence of the MOT is monitored using a CCD camera. An electrode is placed on the top viewport of the vacuum chamber to control the electric field.}
\label{fig:level_scheme_setup}
\end{figure}

Rydberg atoms can be produced in the MOT by a two photon excitation scheme, as illustrated in figure \ref{fig:level_scheme_setup}\,a). The first photon with a wavelength of $\lambda_{MOT} \approx 399\ein{nm}$ corresponds to the $6s^{2}$ $^1S_0$  $\rightarrow$ $6s6p$ $^1P_1$ MOT transition and is provided by the MOT laser. To transfer the atoms from the $6s6p\,^1P_1$\,- state into a Rydberg state of the $6sns\,^1S_0$\,- or the $6snp\,^1P_1$\,- series, the MOT is illuminated with an additional so-called Rydberg laser with a wavelength of $\lambda _R \approx 394\ein{nm}$.  The nominally forbidden transitions $6s6p\,^1P_1 \rightarrow 6snp\,^1P_1$ can be excited, if an external electric field is applied which results in an admixing of even-parity states to the odd-parity $6snp$ $^1P_1$ states. The Rydberg laser light is generated using a commercial diode laser system (Toptica DLpro) which is frequency doubled to $\lambda_{R}$, using a LBO crystal integrated in a home-built bow-tie resonator. The output of the doubling resonator is delivered to the experiment using an optical fiber and is focused onto the MOT with a beam waist of $(100 \pm 10) \ein{\mu m}$ and a power of typically $1\ein{mW}$. The wavelength of the Rydberg laser is measured using a home-built Michelson wavemeter, which is referenced to a rubidium spectroscopy on a daily basis. The reference laser for the wavemeter itself is provided by the master laser for Yb cooling. Like the MOT and the slower light, the Rydberg laser can be switched using mechanical shutters and acousto-optical modulators.\\

A top view of the experimental setup is shown in Fig.\,\ref{fig:level_scheme_setup}. The MOT is located in a cylinder-shaped vacuum chamber with an inner diameter of $100\ein{mm}$ operating at a base pressure below $10^{-10}\ein{mbar}$. In the horizontal plane, the vacuum chamber possesses eight flange connections, two of which are used for the atomic beam and the counter-propagating slower laser beam, four for the horizontal MOT beams and two for imaging purposes. The vertical MOT beams are sent into the chamber through two additional viewports. The MOT magnetic field is generated by two water-cooled solenoids of different size which are mounted outside the vacuum chamber. The smaller coil with a diameter of  $25\ein{mm}$ is mounted in a recessed flange on the bottom of the chamber while the larger coil with a diameter of  $155\ein{mm}$ is mounted on top at a distance of $76\ein{mm}$ from the smaller one. In this configuration, the MOT is positioned very close to the smaller coil and the (grounded) walls of the vacuum chamber nearby. The Zeeman-slowed atomic beam which loads the MOT is generated by a single nozzle oven heated up to $420 \ein{^\circ C}$. Under typical experimental conditions, $10^7$ atoms are captured in the MOT at a temperature of several $\ein{mK}$. The peak density in the MOT is on the order of $\rho_{max}\approx 10^9 \ein{cm^{-3}}$.\\

A relatively homogeneous electric field at the MOT position can be created by simply placing an electrode on the top viewport of the grounded vacuum chamber at a distance of $75 \ein{mm}$ from the MOT position. The electrode is a copper disk with a diameter of $150 \ein{mm}$ and a $25 \ein{mm}$ bore in the center which is used for the vertical MOT beam. Typically, a voltage of a few $\ein{V}$ is applied to the electrode leading to electric fields up to a few $10\ein{V/m}$ at the position of the MOT.\\

\section{Measurement Procedure}
\label{sec:measurement}
A cycle of the sequence, we use for the spectroscopic determination of the excitation energy of Rydberg levels of $^{174}$Yb, is illustrated in Fig.\,\ref{fig:sequence} a). The cycle is initialized by turning off the slower and  MOT light for $100 \ein{ms}$ to remove all remaining atoms from the previous cycle. In the loading phase, the slower and the MOT light are switched on with the MOT light detuned by $\approx 22\ein{MHz}$ from the atomic resonance. The MOT is loaded for $3.5 \ein{s}$, to ensure saturation of the number of atoms in the MOT. Subsequently, the slower light is switched off and the Rydberg laser is switched on during the spectroscopy phase. If the Rydberg laser is resonant with a transition from the $6s6p\, ^1P_1$ state to a Rydberg state, atoms are removed from the MOT and correspondingly the MOT fluorescence is reduced. To determine the relative loss induced by the Rydberg laser, fluorescence images of the MOT are taken right before the Rydberg laser is turned on and after $1 \ein{s}$ of illumination. During this time, the frequency of the Rydberg laser is ramped over $10\ein{MHz}$. The whole cycle is repeated between 50 and 100 times with varying frequency of the Rydberg laser in order to take  a spectrum over a specific frequency range. Every fifth cycle is a calibration cycle, where the Rydberg laser is not switched on in the spectroscopy phase. This calibration is required to account for long-time fluctuations of the number of atoms trapped in the MOT.

\begin{figure}[h]
	\includegraphics[width=\textwidth]{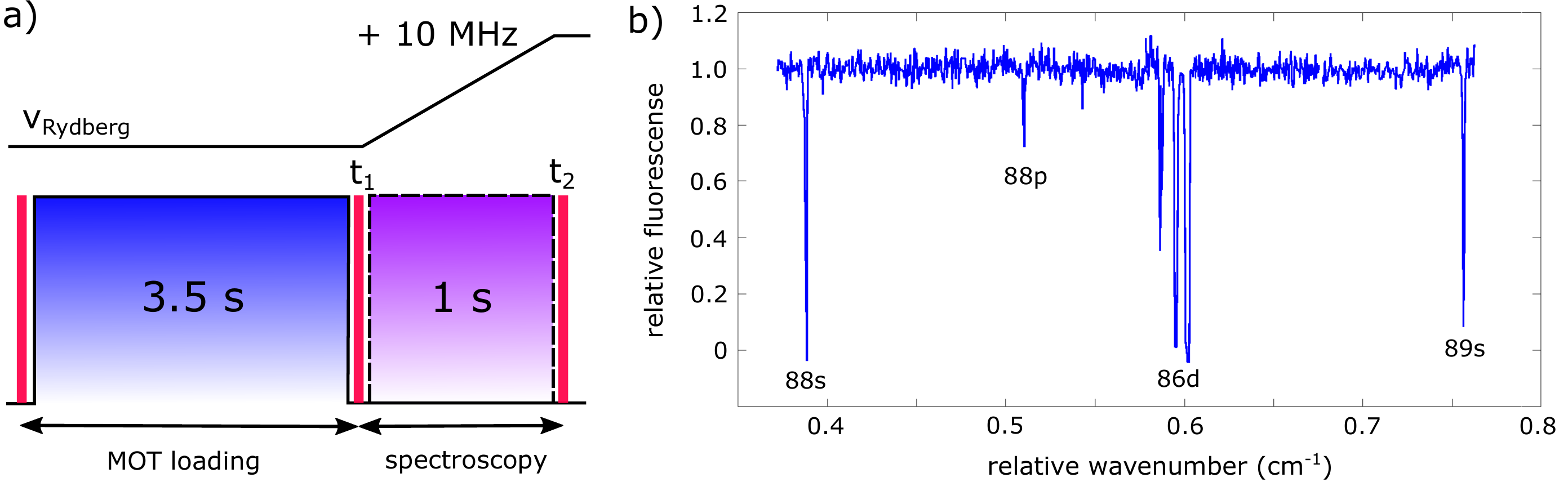}
	\caption{a) Experimental cycle for trap-loss spectroscopy. Each cycle is divided into an initialization phase, a MOT loading phase and a spectroscopy phase. The cycle is repeated between 50 and 100 times to record a spectrum. \\
		b) Typical trap-loss spectrum of $^{174}$Yb Rydberg states. Besides the $6s88s\,^1S_0$-, $6s89s\,^1S_0$- and the $6s88p\,^1P_1$-state also D-levels can be seen in the spectrum. The offset of the $x$-axis is $50427 \ein{cm^{-1}}$.}
	\label{fig:sequence}
\end{figure}

A typical trap-loss spectrum is depicted in Fig.\,\ref{fig:sequence} b). The excitation frequencies (given in wave numbers) are calculated by summing the excitation frequency of the intermediate $6s6p\,^1P_1$ state and the measured frequency of the Rydberg laser which has an error of $\approx 0.005 \ein{cm^{-1}}$ due to the uncertainty of the wavemeter calibration. Other sources of error such as the Zeeman shift due to the MOT magnetic field are estimated to be significantly smaller and are thus not taken into account.

\begin{figure}[h]
	\includegraphics[width=\textwidth]{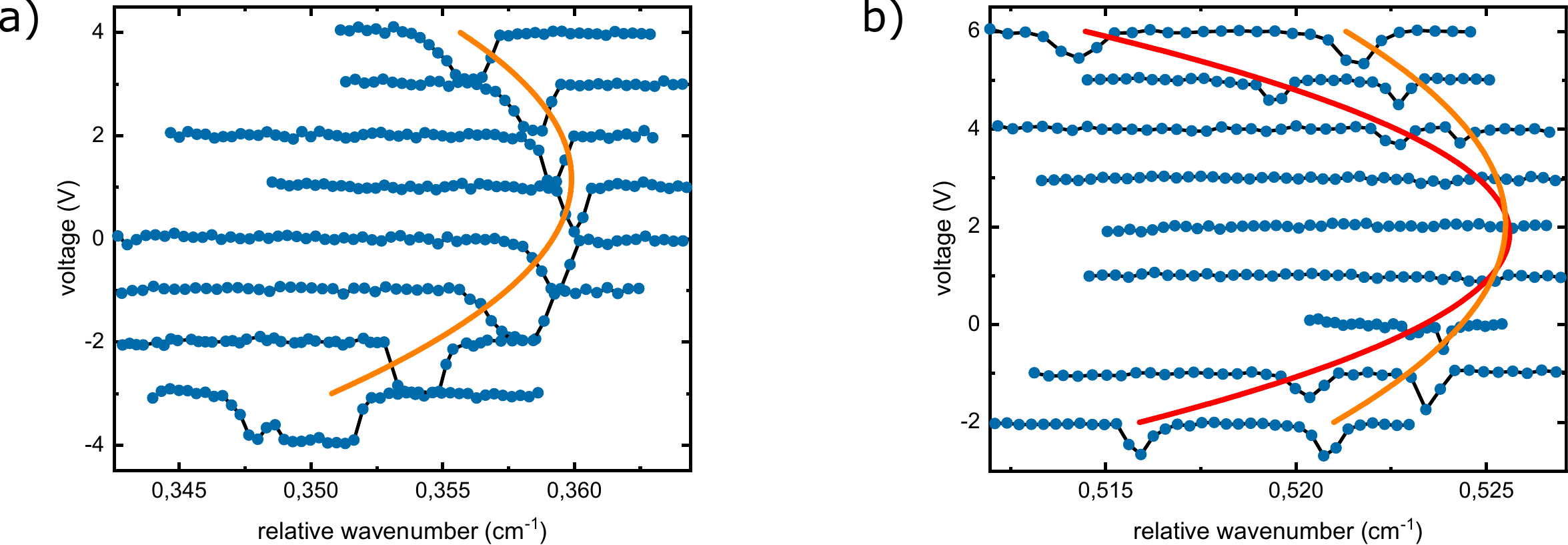}
	\caption{Line spectra of the $6s83s\,^1S_0$ (a) and the $6s83p\,^1P_1$ state (b) for various values of an externally applied electric field. The solid lines are quadratic fits to the Stark shifted line centers corresponding to Eq.\,\ref{eq:Stark_energy}. Due to a non-zero electric background field the apex of the fits is shifted with respect to the zero value of the external electric field. In the case of the $6s83p\,^1P_1$ state, the Rydberg line disappears at zero field due to selection rules. At non-vanishing electric field, it splits into two lines corresponding to $|m_J| = 0$ and $|m_J|=1$. The $x-$axis has an offset of $50425 \ein{cm^{-1}}$ for both graphs.}
	\label{fig:starkmaps}
\end{figure}

To determine the polarizability of a specific Rydberg state, line spectra are taken at various values of an externally applied electric field $\vec{F}$. The external field is generated by applying a voltage to the electric field plate which is located on top of the vacuum chamber. Due to the complex geometry of our chamber and the uncertainty in the exact position of the MOT relative to grounded surfaces the  corresponding electric field has a large uncertainty which is estimated to be more than $50\%$. Therefore, we only give the value of the externally applied voltage in Fig.\,\ref{fig:sequence}. Nevertheless, the experimental geometry ensures that the electric field $\vec{F}$ points predominantly in the vertical direction at the position of the atoms.\\
	
In Fig.\,\ref{fig:starkmaps}, typical spectra are shown for the $6s83s\,^1S_0$ state (a) and the $6s83p\,^1P_1$ state (b), where spectra for different values of the applied voltage are combined into a single graph and the maxima are fitted to a function which is quadratic in the applied voltage (see Eq.\,\ref{eq:Stark_energy}). For the relatively small values of the  applied voltage, only Rydberg states experience a noticeable electric-field (Stark) shift and hence an observed line shift corresponds directly to an energy shift of the Rydberg state. Hence, we  will refer to plots as shown in Fig.\,\ref{fig:starkmaps} as Stark maps. \\

A notable feature of the obtained Stark maps is that the $6s83s^1S_0$ line is only shifted by the external electric field while the $6s83p\,^1P_1$ line is additionally split into two components. This reflects the fact that the Stark shift depends on the absolute value $|m_J|$ of the magnetic quantum number (see Eq.\,\ref{eq:polarizability_m_states}) which can take the values $|m_J| = 0,1$ for the $6s83p\,^1P_1$ state while for the  $6s83s\,^1S_0$ state only $|m_J|=0$ is possible. At the apex of the  $6s83p\,^1P_1$ Stark map, we observe almost no excitation of the atoms into the Rydberg state, in agreement with selection rules that forbid transitions between the two odd-parity states, $6s6p\,^1P_1$ and $6s83p\,^1P_1$. At finite electric field, even-parity states are admixed to the $6s83p\,^1P_1$ state and transitions become possible.

In addition, the apex of the Stark maps is not at zero external field and is not even at an identical position for different Stark maps. We attribute this to background electric fields in the chamber which are most likely originating from patch charges on the top window. Those patch charges may vary on the time scale of days. To remove these charges, at least partially, we regularly illuminate the top window with a $340 \ein{nm}$ LED \cite{Pollack2010} in between experimental runs.  Vanishing of the $6s83p\,^1P_1$\,- line at the apex of the Stark map indicates that the applied external electric field which is oriented vertically compensates the background field to a large extent and we may assume that the background field is also oriented vertically.

 \section{Results and Discussion}

Our results on the excitation energies and polarizabilities of the $6sns\,^1S_0$ and the $6snp\,^1P_1$ series of $^{174}$Yb in the range from $n=70$ to $n=90$ are summarized in table \ref{tab:data}. For each line a complete Stark map was recorded to determine the polarizability and the zero-field excitation energy. The excitation energies for the $6snp\,^1P_1$ series are extrapolated from quadratic fits to the Stark maps (see Fig. \ref{fig:starkmaps}). 
\begin{table}[h]
		\begin{tabular}{c | c | c | | c | c  |c c | c c}
		\hline 
		State & $E_{exc}$ & $\alpha_0$ & State & $E_{exp}$ & $\alpha_a$ & $\alpha_b$ & $\alpha_0$ & $\alpha_2$  \\ 
		\hline 
		$6s70s^1S_0$ & 50417.693 &  1.00 & $6s70p^1P_1$ & 50417.915 & 0.85 & 0.38 & 0,54 & -0,16\\ 
		
		$6s71s^1S_0$ & 50418.420 &  0.88 & $6s71p^1P_1$ & 50418.661 & 1.00 & 0.37 & 0,58 & -0,21\\ 
		
		$6s72s^1S_0$ & 50419.142 &  1.02 & $6s72p^1P_1$ & 50419.372 & 1.00 & 0.43 & 0,62 & -0,19\\ 
		
		$6s73s^1S_0$ & 50419.835 &  1.18 & $6s73p^1P_1$ & 50420.051 & 1.59 & 0.56 & 0.90 & -.34\\ 
		
		$6s74s^1S_0$ & 50420.497 &  1.37 & $6s74p^1P_1$ & 50420.705 & 1.20& 0.54 & 0.76 & -0.22\\ 
		
		$6s75s^1S_0$ & 50421.136 &  1.80 & $6s75p^1P_1$ & 50421.328 & 1.27 & 0.42 & 0.70 & -.28 \\ 
		
		$6s76s^1S_0$ & 50421.737 &  1.77 & $6s76p^1P_1$ & 50421.928 & 1.59  & 0.56 & 0.90 & -0.34 \\ 
		
		$6s77s^1S_0$ & 50422.326 &  1.55 & $6s77p^1P_1$ & 50422.505 & 1.60 & 0.65 & 0.97 & -0.32\\ 
		
		$6s78s^1S_0$ & 50422.881 &  1.53 & $6s78p^1P_1$ & 50423.058 & 1.78 & 0.63 & 1.01 & -0.38\\ 
		
		$6s79s^1S_0$ & 50423.422 &  2.09 & $6s79p^1P_1$ & 50423.588 & 2.10 & 0.72 & 1.64 & -0.46\\ 
		
		$6s80s^1S_0$ & 50423.935 &  2,83 & $6s80p^1P_1$ & 50424.095 & 2.19 & 0.76 & 1.24 & -0.48\\ 
		
		$6s81s^1S_0$ & 50424.432 &  2.45 & $6s81p^1P_1$ & 50424.585 & 2.65 & 0.79 & 1.41 & -0.62\\ 
		
		$6s82s^1S_0$ & 50424.907 &  2.64 & $6s82p^1P_1$ & 50425.053 & 3.05 & 1.01 & 1.69 & -0.68\\ 
		
		$6s83s^1S_0$ & 50425.365 &  2.84 & $6s83p^1P_1$ & 50425.510 & 3.13 & 1.03 & 1.73 & -0.70\\ 
		
		$6s84s^1S_0$ & 50425.805 &  3.54 & $6s84p^1P_1$ & 50425.945 & 3.22 & 1.02 & 1.75 & -0.73\\ 
		
		$6s85s^1S_0$ & 50426.228 &  3.17 & $6s85p^1P_1$ & 50426.364 & 3.69 & 1.45 & 2.20 & -0.75\\ 
		
		$6s86s^1S_0$ & 50426.638 &  3.70 & $6s86p^1P_1$ & 50426.765 & 3.78 & 1.51 & 2.02 & -0.76\\ 
		
		$6s87s^1S_0$ & 50427.031 &  3.92 & $6s87p^1P_1$ & 50427.160 & 4.41 & 1.45 & 2.43 & -0.99\\ 
		
		$6s88s^1S_0$ & 50427.413 &  4.92 & $6s88p^1P_1$ & 50427.532 & 4.41 & 1.57 & 2.52 & -0,95\\ 
		
		$6s89s^1S_0$ & 50427.781 &  4.69 & $6s89p^1P_1$ & 50427.898 & 4.67 & 1.56 & 2.60 & -1.04\\ 
		
		$6s90s^1S_0$ & 50428.137 &  5.60 & $6s90p^1P_1$ & 50428.244 & 3.80 & 1.07 & 1.98 & -0.91\\ 
		\hline
		\hline 
	\end{tabular}
	\caption{Experimental excitation energies of the $6sns\,^1S_0$ and the $6sns\,^1P_1$ Rydberg series of $^{174}$Yb and polarizabilities relative to the measured polarizability of the $6s70s\,^1S_0$ state. The stated values for $E_{exc}$ are the measured energy differences between the $6s^2\,^1S_0$ ground state and the respective Rydberg state in $\ein{cm^{-1}}$. For the $6sns\,^1S_0$ series, $\alpha_0$ is the experimentally determined relative scalar polarizability. For the $6snp\,^1P_1$ series $\alpha_a$  and $\alpha_b$ are the experimentally determined relative polarizabilities corresponding to $|m_J|=0,1$ and $\alpha_0$ and $\alpha_2$ are the derived values of the scalar and tensor polarizabilities according to Eq.\,\ref{eq:polarizability_m_states}. The accuracy of the excitation energies is limited by the accuracy of our home-built wavemeter and is estimated to be $150 \ein{MHz}$ or correspondingly $0.005 \ein{cm^{-1}}$. For the measured polarizabilities the estimated statistical error is $10 \%$. }
	\label{tab:data}
\end{table}

In a system with two valence electrons like Yb a full theoretical description of the Rydberg series can be obtained using Multi-Channel Quantum Defect Theory \cite{Fano1975, Lehec2018}. However, for large principal quantum numbers and in the absence of perturbations by doubly excited states, the excitation energies of a Rydberg series are to a good approximation described by a single quantum defect according to
\begin{eqnarray}
	E_{exc} = E_{ion} - \frac{Ry}{(n-\delta_{L,J})^{2}} =  E_{ion} - \frac{Ry}{n*^{2}}\ .
	\label{eq:excitaion_energies}
\end{eqnarray}  
Here $Ry=10973696.959\ein{m^{-1}}$ is the mass corrected Rydberg constant and  $E_{ion} = 50443.07041(25) \ein{cm^{-1}}$ the first ionization energy \cite{Lehec2018}. The effective principal quantum number $n* = n- \delta_{L,J}$ is determined by the quantum defect $\delta_{L,J}$, which describes the deviation from the hydrogenic case. It is due to the incomplete shielding of the charge of the nucleus by the inner electrons, which is particularly important for small $L$ \cite{Gallagher2004}.

\begin{figure}[h]
\includegraphics[width=\textwidth]{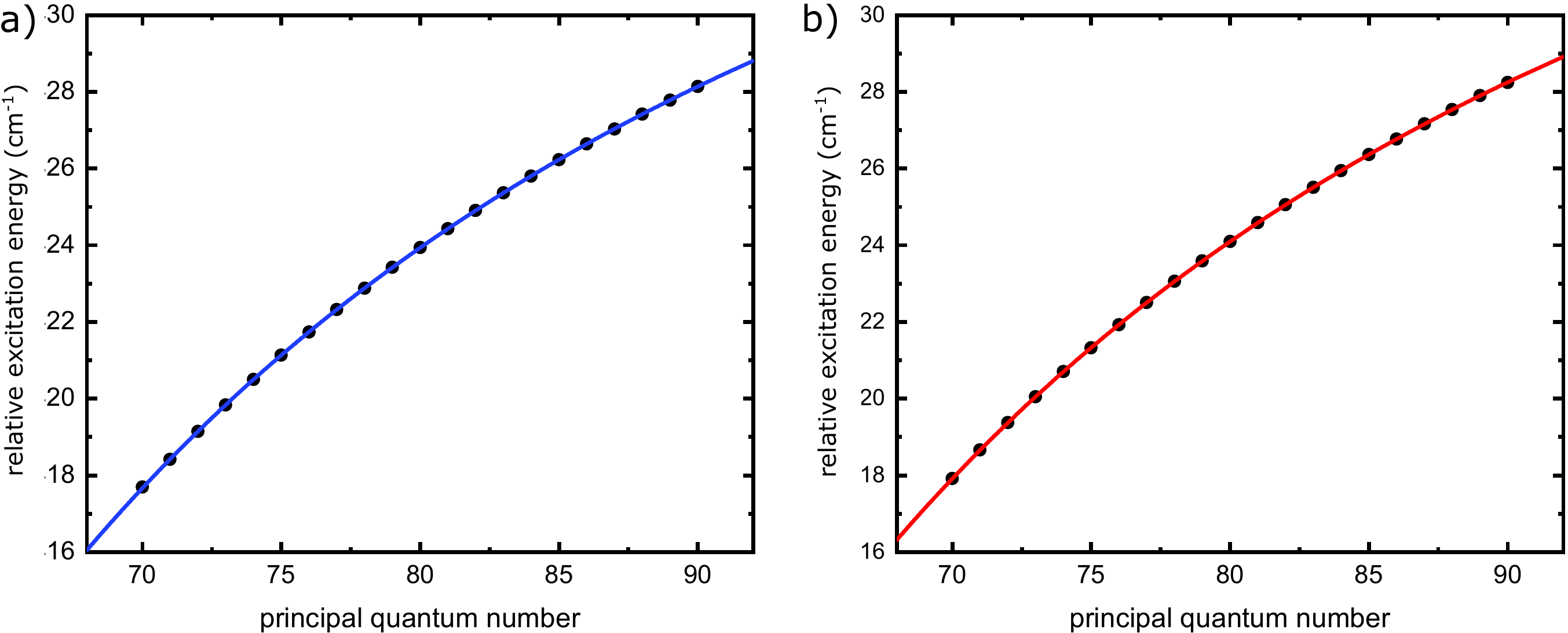}

\caption{Excitation energies for the $6sns\,^1S_0$ (a) and the $6snp\,^1P_1$ (b) Rydberg-series of $^{174}$Yb. The solid lines correspond to fits according to Eq. \ref{eq:excitaion_energies} with $\delta_{L,J}$ being the only free fitting parameter. The obtained values for the quantum defect are $\delta_{0,0} = 4.2721(25)$ and $\delta_{1,1} = 3.9535(9)$. The $y-$axis has an offset of $50400 \ein{cm^{-1}}$.}
\label{fig:state_energies}
\end{figure}

In Fig.\,\ref{fig:state_energies}, the measured excitation energies are plotted as a function of the principle quantum number. The values for the $6sns\,^1S_0$ series that we have obtained using the simple trap-loss method are in good agreement with previously reported values by the group at Laboratoire Aim\'{e}-Cotton \cite{Lehec2018} that have been measured using field-ionization of Rydberg states. From fits of the measured excitation energies to Eq.\,\ref{eq:excitaion_energies} we get the quantum defects $\delta_{0,0} = 4.2721(25)$ and $\delta_{1,1} = 3.9535(9)$, where the value for the $6sns\,^1S_0$ series is again in good agreement with Ref.\,\cite{Lehec2018}. 

If Rydberg spectroscopy is performed in a small external electric field $\vec{F}$ a quadratic Stark shift
\begin{eqnarray}
	\Delta E = -\frac{1}{2}\alpha |\vec{F}|^{2}
\end{eqnarray} 
has to be taken into account where $\alpha$ is the state-dependent atomic polarizability. The total excitation energy is then given by
\begin{eqnarray}
		E_{exc}(F) = E_{ion} - \frac{Ry}{n*^{2}}-\frac{1}{2}\alpha |\vec{F}|^{2} \ . 
		\label{eq:Stark_energy}
\end{eqnarray}
\begin{figure}[h]
	\includegraphics[width=\textwidth]{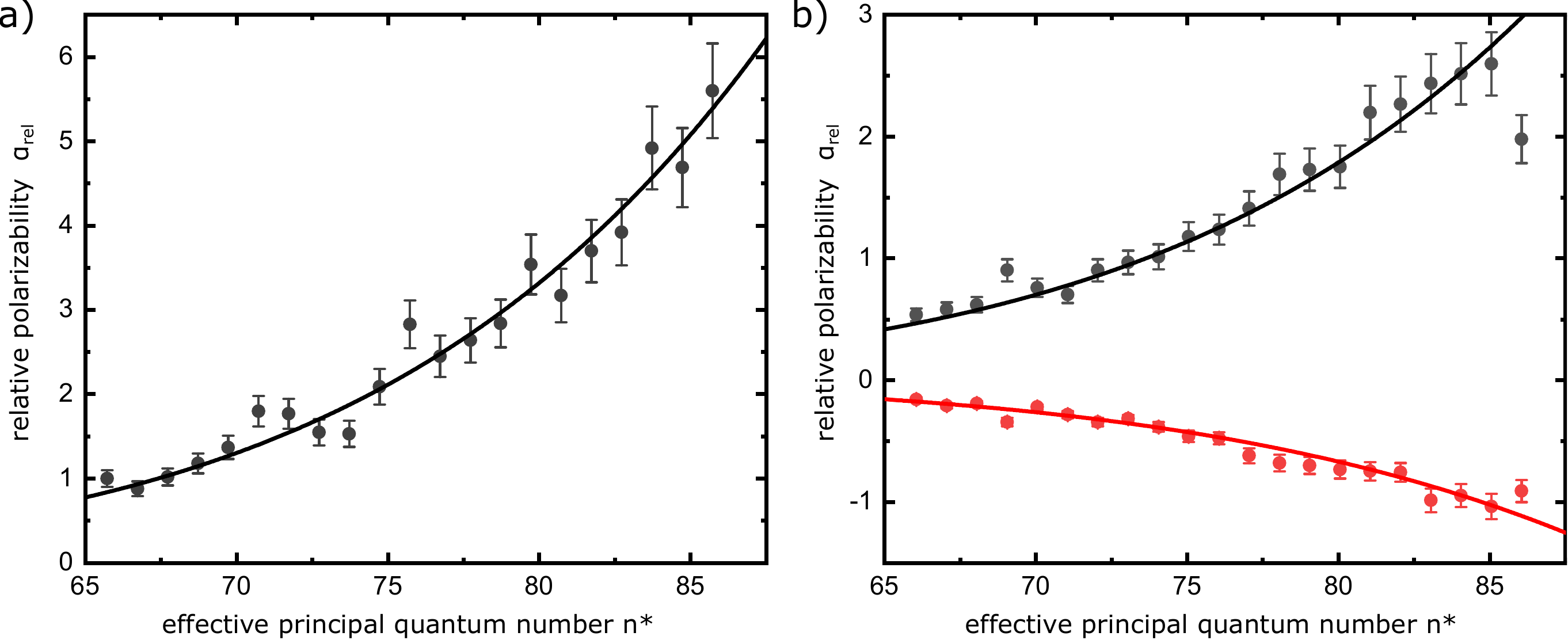}
	\caption{Relative scalar and tensor polarizabilities of the $6sns\,^1S_0$ (a) and the $6snp\,^1P_1$ (b) Rydberg-series of $^{174}$Yb. The scalar polarizability $\alpha_0(n,0)$ of a state $6sns\,^1S_0$ is directly determined from a quadratic fit to the external electric field (see Fig. \ref{fig:starkmaps} a)). The Scalar polarizabilities $\alpha_0(n,1)$ (black dots) and the tensor polarizabilities  and $\alpha_2(n,1)$ (red dots) of the $6snp\,^1P_1$ states are determined using quadratic fits and making use of Eq.\,\ref{eq:polarizability_m_states}. All polarizabilities are normalized to the measured scalar polarizability of the $6s70s\,^1S_0$ state. Solid lines are fits of the polarizabilities to functions of the type $(a(n-\delta_{L,J}))^7 = (a\, n*)^7 $.} 
\label{fig:rel_pol_new}	
\end{figure}

The polarizability $\alpha$ of a Rydberg state with principal quantum number $n$, total angular momentum $J$ and magnetic quantum number $m_J$ can generally be written as \cite{Mitroy2010} 
\begin{eqnarray}
	\alpha (n,J, |m_J|) =\alpha_{0}(n,J)+\alpha_{2}(n,J)\frac{3m_J^{2}-J(J+1)}{J(2J-1)}\,,
	\label{eq:polarizability_m_states}
\end{eqnarray}
with the scalar polarizability $\alpha_{0}(n,J)$ and the tensor polarizability $\alpha_{2}(n,J)$. Thus, the external electric field splits a Rydberg line into $J+1$ components which can be assigned to different absolute values $|m_J|$. 
For the $6sns\,^1S_0$ series, $J=0$ and thus no splitting is observed. The polarizability which is determined as described in Sec.\,\ref{sec:measurement} then corresponds directly to the scalar polarizability $\alpha_0(n, 0)$. For the $6sns\,^1P_1$ series with $J=1$, scalar and tensor polarizability may be derived from the experimentally determined values $\alpha(n,1,1)$ and $\alpha(n,1,0)$ of the two observed components (see Fig.\,\ref{fig:starkmaps}\,b) according to 
\begin{eqnarray}
	\alpha_0(n,1) &=& \frac13 \left( 2 \alpha(n,1,1) + \alpha(n,1,0)\right)\ \ \mathrm{ and }\nonumber\\
	\alpha_2(n,1) &=& \frac13 \left(\alpha(n,1,1) - \alpha(n,1,0))\right)\,. 
\end{eqnarray}
Using these formulas, the values for $\alpha_0(n,1)$ and $\alpha_2(n,1)$ given in table \ref{tab:data} are calculated under the assumption that the tensor polarizability $\alpha_2(n,1)$ is negative implying that the measured component with the larger polarizability $\alpha_a (n,1)$ for a given state of the $6sns\,^1P_1$ series corresponds to $J=1, |m_J|=0$. This assumption seems justified as it resembles the behavior of the polarizabilities in highly excited $^2P_{3/2}$ Rydberg states of rubidium \cite{Yerokhin2016,Lai2018}.
In Fig.\,\ref{fig:rel_pol_new}, the determined scalar and tensor polarizabilities for all states of the $6sns\,^1S_0$ (a) and the $6snp\,^1P_1$ series of $^{174}$Yb in the range of $n=70$ to $n=90$ are shown. The solid lines in the figure are fits of the polarizabilities to functions of the type $(a(n-\delta_{L,J}))^7 = (a\, n*)^7 $, which reflect the expected scaling with $n*^7$ of the polarizabilities of highly excited Rydberg states \cite{Gallagher2004}.  Due to the large uncertainty of the value of the external electric field at the position of the atoms all polarizabilities are normalized to the measured polarizability of the $6s70s\,^1S_0$ state.

\section{Conclusion}

In this manuscript, we have presented a spectroscopic measurement of the $6sns\,^1S_0$ and the $6snp\,^1P_1$ Rydberg-series of $^{174}$Yb which is based on simple trap-loss spectroscopy in a magneto-optical trap. Using this method, we have obtained excitation energies for Rydberg states in the range of $n=70$ to $n=90$ which are in agreement with previously reported results on the $6sns\,^1S_0$-series and extend those results to larger principal quantum numbers and the $6snp\,^1P_1$ series. In addition, we have performed an investigation of the Stark effect in low electric fields allowing us to infer scalar and tensor polarizabilities which exhibit the expected $n*^7$ scaling behavior. These results will be valuable for further experiments using Yb Rydberg atoms.

\section*{Data availability}

Research data supporting this publication are available from the corresponding author upon reasonable request.
\newpage
%\bibliographystyle{unsrt}
%\bibliography{References_General}

%\printbibliography

\end{document}